\documentstyle[12pt,fps]{article}
\newcommand{\Z}{Z \!\!\! Z}
\newcommand{\halb}{1/2}
\newcommand{\D}{{\cal D}}
\makeatletter
\@addtoreset{equation}{section}
\makeatother

\begin{document}
\vspace*{1cm}
\begin{center}
{\Large Perfect Lattice Actions for Staggered Fermions}
\footnote{This work is supported in part by funds provided
by the U.S. Department of Energy (D.O.E.) under
cooperative research agreement DE-FC02-94ER40818.}

\vspace*{1cm}

W. Bietenholz$^{\rm a}$, R. Brower$^{\rm b}$, 
S. Chandrasekharan$^{\rm c}$ and U.-J. Wiese$^{\rm c}$

\vspace*{1cm}

$^{\rm a}$ HLRZ c/o KFA J\"{u}lich \\
52425 J\"{u}lich, Germany \\
\ \\

$^{\rm b}$ Department of Physics \\
Boston University \\
Boston MA 02215, USA \\ 
\ \\

$^{\rm c}$ Center for Theoretical Physics \\
Laboratory for Nuclear Science and Department of Physics \\
Massachusetts Institute of Technology \\
Cambridge MA 02139, USA \\
\ \\

Preprint MIT-CTP 2584, HLRZ 74/96

\end{center}

\vspace*{15mm}

We construct a perfect lattice action
for staggered fermions by blocking from the continuum.
The locality, spectrum and pressure of such
perfect staggered fermions are discussed.
We also derive a consistent fixed point action for
free gauge fields and discuss its locality as well as
the resulting static quark-antiquark potential. 
This provides a basis for the construction of 
(classically) perfect lattice actions for
QCD using staggered fermions.

\newpage

\section{Introduction}

Recently there has been a surge of interest in lattice actions
which are closer to the continuum limit than standard lattice
actions, in the sense that artifacts due to the finite
lattice spacing are suppressed. The hope is that such improved
actions will allow Monte Carlo simulations to reach the scaling
region even on rather coarse lattices.

Based on renormalization group concepts, it has been known
for a long
time that perfect actions, i.e. lattice actions without any
cutoff artifacts, do exist \cite{WilKog}.
However, it is very difficult to construct -- or
%at least (Rich,Sch)
even approximate -- such perfect actions.
%A recent (Rich)
Recent progress 
%was (Rich)
is based on the observation that for asymptotically
free theories the determination of the fixed point action (FPA)
is a classical field theory problem \cite{HasNie}.
At infinite correlation length, a FPA is perfect.
Away from the critical surface, this is in general
not the case
any more; at finite correlation length, FPAs are
referred to as ``classically perfect actions''.
However, they are considered
as promising improved actions even at moderate 
correlation length. In particular, the deviations from
the (quantum) perfect renormalized trajectory are likely
to set in only at the two loop level \cite{Hung,Mafia}.
Indeed, a drastically improved scaling behavior of the
fixed point action has been observed in some asymptotically
free models \cite{HasNie,GN,Hung}, and is especially
hoped for in QCD.

A classically perfect action can be constructed as a fixed point
of a block variable
renormalization group transformation (RGT). Usually one fixes
a finite blocking factor $n$ and builds block variables on a coarse
lattice of spacing $n$, determined by a fine lattice with unit
spacing. Then one expresses all quantities in the new lattice
units, i.e. one rescales the coarse lattice back to unit spacing.
At infinite correlation length and for suitable RGT parameters,
an infinite number of iterations of this RGT step leads to a
finite FPA. 
%As we mentioned before,
%the FPA is perfect at $\xi =\infty $,
%and classically perfect at finite $\xi$.
For most interacting theories, this scheme is the only way to
construct 
%FPAs (Rich)
FPA's non-perturbatively. There one has to perform the
blocking transformation numerically 
(and hope for 
%a (Rich)
swift convergence).

For free or perturbatively interacting theories this construction
can also be achieved in only one step, by a technique that we
call ``blocking from the continuum''. 
This method 
% runs (Uwe)
proceeds
analytically in momentum space.
It has been applied to construct
perturbatively perfect lattice actions for the Schwinger model and
for QCD, using fixed point fermions of the Wilson type \cite{Schwing,QuaGlu}.
In this paper we present a number of ingredients for the construction
of a classically perfect QCD lattice action with staggered fermions,
using the same technique.
This provides a construction scheme alternative to Ref. \cite{QuaGlu}.
A motivation is that staggered fermions are particularly useful for the 
study of chiral symmetry breaking.

In section 2 we derive the perfect action for free staggered
fermions by blocking from the continuum. In section 3 we discuss
the dispersion relation and the pressure of such fixed point fermions
with and without truncation. In section 4 we show how to block
the gauge field consistently and we arrive at the
corresponding FPA for free gauge fields.
This implies a ``classically perfect'' static quark-antiquark
potential, which is compared to the (perfect) Coulomb
potential in section 5.
%To provide a basis for the construction of
%non-perturbative FPAs -- which requires
%numerical RGTs -- we relate these results to RGTs with finite
%blocking factors in section 6. 
Section 6 contains our conclusions and an outlook on possible
applications.
%, regarding in particular the pion mass splitting.

A synopsis of the results presented here is included
in Ref. \cite{StL}, where we also illustrate
the perturbative blocking from the continuum and where we outline
how perturbatively perfect actions can serve as
starting points for the non-perturbative search for 
%FPAs.
FPA's.

\section{Perfect free staggered fermions}

The fixed point action for free, massless staggered
fermions in two dimensions
has been derived before \cite{Dallas,GN}.
That derivation used a block factor 3 RGT, which does
not mix the pseudoflavors \cite{Kalk}, and which could be iterated
analytically an infinite number of times.
We are going to show how one can reproduce this result in
one step by blocking from the continuum. 
\footnote{This has also been achieved by H. Dilger, in a way
which emphasizes the relationship to Dirac-K\"{a}hler fermions
\cite{Dilg}. A similar suggestion occurred earlier in Ref. \cite{Hambu}.
As a further approach one could first construct perfect naiv
fermions, including all the doublers, and then build perfect staggered
fermions from them. We thank M.-P. Lombardo for this remark.}
We also generalize
the result to higher dimensions and
to arbitrary masses, so we obtain an entire
renormalized trajectory for free fermions.

First we introduce our notation for staggered fermions.
We divide the lattice into disjoint hypercubes with centers $x$.
Our lattice has spacing 1/2, such that these centers have spacing 1.
Each hypercube carries a set of $2^{d}$ fermionic and
antifermionic pseudoflavors on its corners, where the dimension
$d$ is assumed to be even. We denote those Grassmann variables as
$\chi_{x}^{\rho}, \ \bar \chi_{x}^{\rho}$, where $\rho$ is the
vector pointing from the cell center $x$ to the corner, where
the variable lives. Hence the standard action for free staggered
fermions reads
\begin{equation}
S_{st}[\bar \chi ,\chi ] = \sum_{x\in \Z^{d}, \rho \rho '}
\bar \chi_{x}^{\rho} \Big[ \sum_{\mu =1}^{d} \Big(
\Gamma_{\mu}^{\rho \rho'} \hat \partial_{\mu} +
\frac{1}{2} \Gamma_{5\mu}^{\rho \rho'} 
\hat \partial^{2}_{\mu} \Big) 
+ 2 m \delta^{\rho \rho'} \Big]
\chi^{\rho'}_{x} ,
\end{equation}
where we have defined the following quantities
\begin{eqnarray} \nonumber
\Gamma^{\rho \rho'}_{\mu} &=& [\delta^{\rho - \hat \mu /2, \rho '}
+ \delta^{\rho+\hat \mu /2, \rho'} ] \sigma_{\mu}(\hat \rho ) , \\
\Gamma^{\rho \rho'}_{5\mu} &=& [\delta^{\rho - \hat \mu /2, \rho '}
- \delta^{\rho+\hat \mu /2, \rho'} ] \sigma_{\mu}(\hat \rho ) , 
\nonumber \\ \nonumber
\sigma_{\mu}(x) &=& \Big\{ \begin{array}{ccc}
\ 1 && \sum_{\nu < \mu} x_{\nu} ~~{\rm even} \\
-1 && {\rm otherwise} \end{array} \\ \nonumber
\hat \partial_{\mu} \chi_{x}^{\rho} &=& \frac{1}{2}
( \chi^{\rho}_{x+\hat \mu} - \chi^{\rho}_{x-\hat \mu}), \\
\hat \partial^{2}_{\mu} \chi_{x}^{\rho} &=&
\chi^{\rho}_{x+\hat \mu} + \chi^{\rho}_{x-\hat \mu}
-2\chi^{\rho}_{x} .
\end{eqnarray}

Now we are going to construct a perfect action for free staggered
fermions. Instead of iterating an RGT with a finite blocking factor,
we send the blocking factor to infinity and perform only
one RGT. This amounts to the ``blocking
from the continuum'': one starts from continuum fields and
defines lattice variables by integrating over the unit hypercube
around the corresponding lattice site. In particular, for staggered
fermions we also have to take care of the flavor structure.
We have to transform the continuum flavors into staggered
pseudoflavors and integrate over regions, which depend on
the pseudoflavor. Those regions are overlapping unit hypercubes.

%We define the lattice staggered fermion fields as
%\begin{eqnarray}
%\chi_{x}^{\rho} &=& \int_{c_{x+ \rho}} d^{d}y \ U^{\rho i}
%\psi^{i} (y) , \\
%\bar \chi_{x}^{\rho} &=& \int_{c_{x + \rho}} d^{d}y \
%\bar \psi^{i} (y) U^{\rho i~\dagger} , \nonumber
%\end{eqnarray}
%where $\psi^{i} , \ \bar \psi^{i} $ are continuum fermion fields
%with flavor index $i$. The unitary matrix $U$ transforms its flavors
%into the staggered pseudoflavors, see e.g. \cite{Smit},
%and $c_{x}$ is a unit hypercube with center $x$.
It is well-known how to transform fermionic pseudoflavors
in the continuum limit into flavors
by a unitary transformation. 
We assume that the inverse of this
transformation has been carried out and we start
in the continuum with space filling fermionic fields
$\psi , \ \bar \psi$, composed of pseudoflavors. Then we build 
staggered lattice fermions as
\begin{equation}
\chi_{x}^{\rho} = \int_{c_{x+ \rho}} d^{d}y \
\psi^{\rho} (y) \ , \quad
\bar \chi_{x}^{\rho} = \int_{c_{x + \rho}} d^{d}y \
\bar \psi^{\rho} (y) \ ,
\end{equation}
where $c_{z}$ is a unit hypercube with center $z$.
We define
\begin{equation}
\Pi (p) = \int_{c_{0}} d^{d}y \exp (ipy) = \prod_{\mu =1}^{d}
\frac{\hat p_{\mu}}{p_{\mu}} \ ; \quad
\hat p_{\mu} = 2 \sin (p_{\mu}/2) ,
\end{equation}
and obtain in momentum space
\begin{eqnarray} \label{fermitrafo}
\chi^{\rho} (p) &=& \sum_{l \in \Z^{d}}
\psi^{\rho}(p+2\pi l) \Pi (p+2\pi l) e^{i(p+2\pi l) \rho}
\nonumber \\
&=& \sum_{l \in \Z^{d}} \psi^{\rho} (p+2\pi l) \Pi^{\rho}
(p+2\pi l) ,
\end{eqnarray}
where
\begin{equation}
\Pi^{\rho}(p) =  e^{ip \rho } \Pi (p)
\end{equation}
and $p$ is in the Brillouin zone $B = ]-\pi ,\pi ]^{d}$.
In analogy to Refs.~\cite{Dallas,GN} we choose the following 
% RGT (Rich)
RGT,
\begin{eqnarray}
e^{-S[\bar \chi ,\chi ]} &=& \int \D \bar \psi \D \psi
\D \bar \eta \D \eta \exp \Big\{ - s[\bar \psi ,\psi ] \nonumber \\
&+& \frac{1}{(2\pi )^{d}} \int_{B} d^{d}p \Big(
[ \bar \chi^{\rho}(-p) - \sum_{l\in \Z^{d}} \bar \psi^{\rho}(-p-2\pi l)
\Pi^{\rho}(-p-2\pi l)] \eta^{\rho}(p) \nonumber \\
&& + \bar \eta^{\rho}(-p) [ \chi^{\rho}(p) - \sum_{l \in \Z^{d}}
\Pi^{\rho}(p+2\pi l) \psi^{\rho} (p+2\pi l) ] \nonumber \\
&& + \bar \eta^{\rho}(-p) A^{\rho \rho'} \eta^{\rho'}(p)
\Big) \Big\}, \label{RGT}
\end{eqnarray}
% (We do not write summation symbols for $\rho , \rho'$ any more). (Sch)
where the summation over $\rho, \rho'$ is understood.
Here $s[\bar \psi ,\psi ]$ is the continuum action and
$\bar \eta^{\rho} , \ \eta^{\rho}$ are auxiliary
staggered Grassmann fields living on the same lattice
sites as $\bar \chi^{\rho}$ and $ \chi^{\rho}$.
We do not enforce the blocking relation (\ref{fermitrafo})
by a 
% $\delta $ function, (Uwe) 
$\delta $-function, 
but only by a smoothly peaked
function; the 
% $\delta $ function (Uwe)
$\delta $-function 
is smeared to a Gaussian by
the matrix $A$. For the latter we make the ansatz,
\begin{equation}
A^{\rho \rho'} = a \delta^{\rho \rho'}
-i c \hat p_{\mu} \Gamma_{\mu}^{\rho \rho'} e^{ip(\rho - \rho ')},
\end{equation}
where $a$ and $c$ are referred to as mass-like and kinetic
smearing parameter. They can be tuned to optimize the
locality of the perfect action. In particular for $a=0$
the RGT preserves the remnant chiral symmetry $U(1) \otimes
U(1)$ (at $m=0$),
which is therefore explicitly present in the perfect action.

Integrating the RGT (\ref{RGT}) we obtain the perfect lattice action
\begin{equation} \label{perfa}
S [\bar \chi  , \chi ] = \frac{1}{(2\pi )^{d}} \int_{B}
d^{d}p \ \bar \chi^{\rho}(-p) [\Delta^{f}(p)^{-1}]^{\rho \rho'}
\chi^{\rho '}(p),
\end{equation}
with the free propagator
\begin{eqnarray}
\Delta^{f}(p)^{\rho \rho '} &=& \sum_{l \in \Z^{d}} \Pi^{\rho}(p+2\pi l)
\frac{1}{i (p_{\mu}+2\pi l_{\mu})\gamma_{\mu}+m}
\Pi^{\rho '}(p+2\pi l) + A^{\rho \rho'} \nonumber \\
&=& -i \alpha_{\mu}(p) \Gamma_{\mu}^{\rho \rho'}
e^{ip(\rho -\rho ')} + \beta(p) \delta^{\rho \rho'} , \label{perfp}
\end{eqnarray}
where $\alpha_{\mu}(p)$ and $\beta (p)$ are given by
\begin{eqnarray} \nonumber
\alpha_{\mu}(p) &=& \sum_{l \in \Z^{d}} \frac{(-1)^{l_{\mu}} (p_{\mu}+
2\pi l_{\mu})}{(p+2\pi l)^{2}+m^{2}} \Pi (p+2\pi l)^{2}
+ c \hat p_{\mu} , \\
\beta (p) &=& \sum_{l \in \Z^{d}} \frac{m}{(p+2\pi l)^{2}+m^{2}} 
\Pi (p+2\pi l)^{2} + a .
\end{eqnarray}
%In the chiral limit this result agrees with the FPA found in
%Ref. \cite{Dallas,GN}.
For $p= (p_{1}, 0 ,\dots ,0)$,
i.e. in the effectively one dimensional case,
the sum over $l$ collapses to a sum over $l_{1}\in \Z$, which
can be computed analytically. In this case, it turns out that for
\begin{equation} \label{spara}
a = \frac{\sinh (m) - m}{m^{2}} \ ; \quad
c = \frac{\cosh (m/2) -1}{m^{2}},
\end{equation}
the action turns into the standard action,
which is {\em ultralocal}; the range of couplings
does not exceed 
%next neighbors. (Rich Sch Uwe)
nearest neighbors.
In higher dimensions the sum over $l$ has to be done
numerically and we can only obtain locality in the
sense of an exponential decay.
It turns out that the same choice of
smearing parameters $a$ and $c$ still yields an extremely
local action, i.e. the exponential decay is extremely fast,
in analogy to our observations for Wilson
fermions \cite{QuaGlu,Dallas}. 
%\footnote{If we apply this method to massless scalar fields,
%we find the optimally local FPA, which was identified
%numerically in Ref. \cite{BeWi}.}
Hence we are going to
use the smearing parameters given in eq.~(\ref{spara})
in any dimension.

In the chiral limit $m = 0$ this result coincides with the
FPA obtained in Refs. \cite{Dallas,GN}.
The optimally tuned value of $a$ vanishes in this limit,
\footnote{This is in contrast to Wilson-like fermions,
where a chiral symmetry breaking smearing parameter is
required for a local FPA \cite{UJW}.}
hence we obtain in the massless case an extremely
local perfect action,
which still has the remnant chiral symmetry.

The inverse propagator can be represented as
\begin{eqnarray}
[\Delta^{f}(p)^{-1}]^{\rho \rho'} &=&
\rho_{\mu}(p) \Gamma^{\rho \rho '}_{\mu} e^{ip(\rho - \rho ')}
+ \lambda (p) \delta^{\rho \rho '} \nonumber \\
\rho_{\mu}(p) &=& \frac{i \alpha_{\mu}(p)}
{\alpha_{\mu}^{2}(p) + \beta^{2}(p)} \ ; \quad
\lambda (p) = \frac{\beta (p)}{\alpha_{\mu}^{2}(p) + \beta^{2}(p)} \ .
\end{eqnarray}
The function $\rho_{\mu}(p)$ is antisymmetric and $2\pi$
antiperiodic in $p_{\mu}$. In all other momentum components it is
symmetric and $2\pi $ periodic, and the same holds for $\lambda (p)$
in all components of $p$. Hence we can expand these functions
in Fourier series,
\begin{equation}
\rho_{\mu}(p) = \sum_{z} \rho_{\mu ,z}
e^{ip z} \ ,
\quad \lambda (p) = \sum_{n \in \Z^{d}} \lambda_{n} e^{ipn} \ ,
\end{equation}
where $z_{\mu} \in \{ \pm 1/2 , \pm 3/2 , \dots \} $ and
$z_{\nu} \in \Z, \ \nu \neq \mu$.
There is a symmetry under permutation and sign flip
of $\lambda_{n}$ in all components of $n$, and of 
$\rho_{\mu ,z}$ in all $z_{\nu}, \ \nu \neq \mu$.
Furthermore $\rho_{\mu ,z}$ is antisymmetric in $z_{\mu}$.

Expressed in these quantities, the staggered fermion action
takes the form
\begin{eqnarray}
S [\bar \chi ,\chi ] &=& \sum_{x\in \Z^{d}} \bar \chi^{\rho}_{x}
\Big\{ \sum_{\mu ,z} \rho_{\mu ,z} \sigma_{\mu} (x)
[ \delta^{\rho - \hat \mu /2,
\rho '} \chi^{\rho '}_{x+z+\hat \mu} +\delta^{\rho + \hat \mu /2,
\rho '} \chi^{\rho '}_{x+z} ] \nonumber \\
&& + \sum_{n} \lambda_{n} \chi^{\rho'}_{x+n} \Big\} .
\end{eqnarray}
%If we rename $\chi^{\rho}_{x} = \chi_{x+\rho}$, i.e. if we
%characterize the fermionic variables solely by their site
%on the lattice of spacing $1/2$, then this expression turns into
%\begin{equation} \label{perfstag}
%S [ \bar \chi ,\chi ] = \sum_{x \in \Z^{d}}
%\bar \chi_{x} \Big\{ \sum_{z} \rho_{\mu ,z} \sigma_{\mu}(x) 
%\chi_{x+z} + \sum_{n} \lambda_{n} \chi_{x+n} \Big\} .
%\end{equation}
The couplings for standard staggered fermions are
\begin{equation} \label{standa}
\rho_{\mu ,z} = ( \delta_{2z_{\mu},1} - \delta_{2z_{\mu},-1} )
\prod_{\nu \neq \mu} \delta_{z_{\nu},0} \ ; \quad
%\lambda_{n} = m \delta_{n,0} .
\lambda_{n} = 2 m\; \delta_{n,0}\; .
\end{equation}
For the perfect action, defined in 
% eqs.~(\ref{perfa}) to (\ref{spara}) (Uwe) Actually Uwe wanted (2.9-12)
% I dont know how to do this!
eqs.~(\ref{perfa})--(\ref{spara}),
the largest couplings
are given in Table 1 and 2 for masses $m=0, \ 1, \ 2$ and 4, 
and their decay is plotted in Fig.~1 and 2. 
We observe an extreme degree
of locality, which is very important for practical purposes.
For numerical application the perfect action has to be
truncated to a short range. This truncation
ought to alter the action as little as possible, in order to
preserve the perfect properties to a good approximation.
\begin{figure}[hbt]
%\vspace{-2mm}
%\begin{center}
\hspace{7mm}
\def\fpsangle{0}
\epsfxsize=120mm
\fpsbox{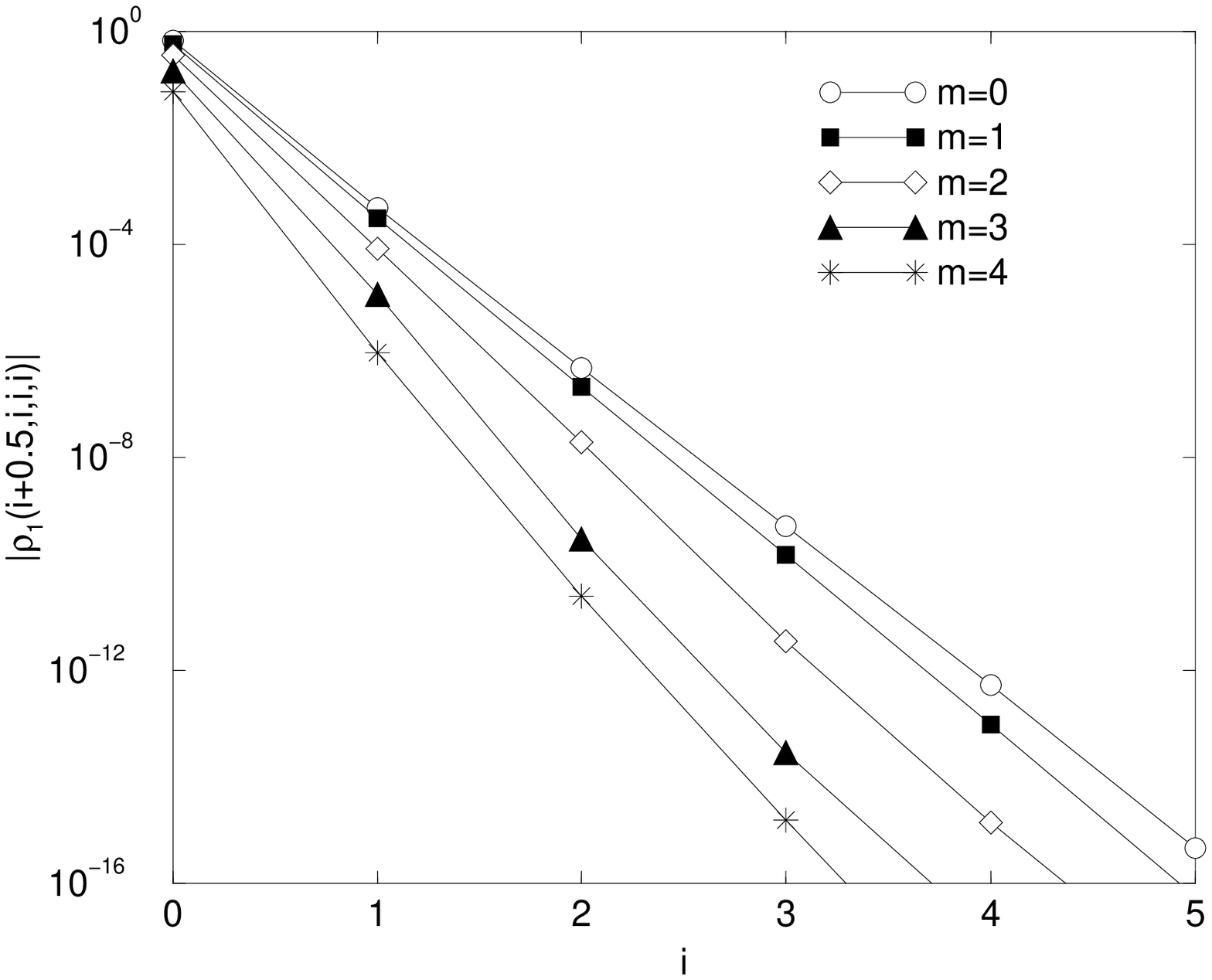}
%\vspace{-5mm}
%\end{center}
\caption{The decay of the kinetic couplings 
$\rho_{1}(i+1/2,i,i,i)$ of the perfect staggered fermion
for masses 0, 1, $\dots$ 4. We observe that the decay is exponential
and very fast. For increasing mass it 
%turns (Rich Sch Uwe)
decays even faster.}
\vspace{-2mm}
\end{figure}

\begin{figure}[hbt]
\hspace{7mm}
\def\fpsangle{0}
\epsfxsize=120mm
\fpsbox{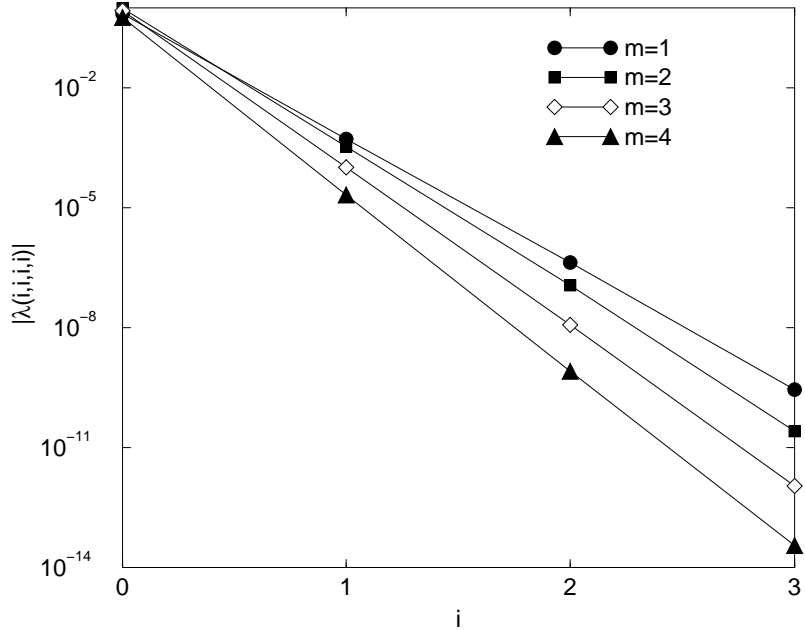}
\caption{The decay of the static couplings $\lambda (i,i,i,i)$
in the perfect staggered fermion action at masses
1, 2, 3 and 4. We confirm that the decay is exponential
and very fast, especially for large masses.}
\vspace{-2mm}
\end{figure}

\begin{table}
\begin{center}
\begin{tabular}{|c|c|c|c|c|}
\hline
$(z_{1},z_{2},z_{3},z_{4})$ & $m=0$ & $m=1$ &$m=2$ &$m=4$ \\
\hline
(\halb ,0,0,0) & ~0.6617391 & ~0.5649324 & ~0.3586038 & ~0.0730572 \\
\hline
(\halb ,0,0,1) & ~0.0441181 & ~0.0335946 & ~0.0154803 & ~0.0011324 \\
\hline
(\halb ,0,1,1) & ~0.0046569 & ~0.0032407 & ~0.0012088 & ~0.0000711 \\
\hline
(\halb ,1,1,1) & ~0.0004839 & ~0.0003251 & ~0.0001282 & ~0.0000114 \\
\hline
(\halb ,0,0,2) & ~0.0018423 & ~0.0012135 & ~0.0003688 & ~0.0000063 \\
\hline
(\halb ,0,1,2) & ~0.0001419 & ~0.0000741 & ~0.0000096 & -0.0000001 \\
\hline
(\halb ,1,1,2) & -0.0000264 & -0.0000197 & -0.0000065 & -0.0000001 \\
\hline
(\halb ,0,2,2) & -0.0000145 & -0.0000117 & -0.0000046 & -0.0000001 \\
\hline
(\halb ,1,2,2) & -0.0000126 & -0.0000080 & -0.0000022 & ~0.0000000 \\
\hline
(\halb ,0,0,3) & ~0.0000780 & ~0.0000445 & ~0.0000090 & ~0.0000000 \\
\hline
(3/2,0,0,0) & ~0.0234887 & ~0.0172839 & ~0.0071677 & ~0.0003303 \\
\hline
(3/2,0,0,1) & -0.0004933 & -0.0005804 & -0.0004502 & -0.0000389 \\
\hline
(3/2,0,1,1) & -0.0009913 & -0.0007019 & -0.0002537 & -0.0000067 \\
\hline
(3/2,1,1,1) & -0.0004819 & -0.0003056 & -0.0000832 & -0.0000009 \\
\hline
(3/2,0,0,2) & -0.0001210 & -0.0000924 & -0.0000366 & -0.0000008 \\
\hline
(3/2,0,1,2) & -0.0001011 & -0.0000623 & -0.0000155 & -0.0000001 \\
\hline
(3/2,1,1,2) & -0.0000462 & -0.0000255 & -0.0000048 & ~0.0000000 \\
\hline
(3/2,0,2,2) & -0.0000129 & -0.0000073 & -0.0000015 & ~0.0000000 \\
\hline
(5/2,0,0,0) & ~0.0009439 & ~0.0006010 & ~0.0001648 & ~0.0000018 \\
\hline
(5/2,0,0,1) & -0.0000600 & -0.0000483 & -0.0000197 & -0.0000003 \\
\hline
(5/2,0,1,1) & -0.0000577 & -0.0000339 & -0.0000070 & ~0.0000000 \\
\hline
(5/2,1,1,1) & -0.0000187 & -0.0000087 & -0.0000007 & ~0.0000000 \\
\hline
(7/2,0,0,0) & ~0.0000390 & ~0.0000215 & ~0.0000039 & ~0.0000000 \\
\hline
\end{tabular}
\end{center}
\caption{The largest kinetic couplings $\rho_{1}(z)$
for the optimally local, perfect staggered fermion
of mass 0, 1, 2 and 4. The table contains all
the couplings $\geq 10^{-5}$.}
\end{table}

\begin{table}
\begin{center}
\begin{tabular}{|c|c|c|c|c|}
\hline
$(z_{1},z_{2},z_{3},z_{4})$ & $m=0$ & $m=1$ &$m=2$ &$m=4$ \\
\hline
(0,0,0,0) & 0 & ~0.7091500 & ~0.9799873 & ~0.5705389 \\
\hline
(0,0,0,1) & 0 & ~0.0296055 & ~0.0246899 & ~0.0034779 \\
\hline
(0,0,1,1) & 0 & -0.0010357 & -0.0015531 & -0.0003674 \\
\hline
(0,1,1,1) & 0 & -0.0012092 & -0.0009711 & -0.0000904 \\
\hline
(1,1,1,1) & 0 & -0.0005181 & -0.0003371 & -0.0000208 \\
\hline
(0,0,0,2) & 0 & ~0.0009257 & ~0.0005581 & ~0.0000181 \\
\hline
(0,0,1,2) & 0 & -0.0000828 & -0.0000703 & -0.0000035 \\
\hline
(0,1,1,2) & 0 & -0.0000567 & -0.0000265 & -0.0000003 \\
\hline
(1,1,1,2) & 0 & -0.0000136 & -0.0000026 & ~0.0000001 \\
\hline
(0,0,0,3) & 0 & ~0.0000332 & ~0.0000133 & ~0.0000001 \\
\hline
\end{tabular}
\end{center}
\caption{The largest static couplings $\lambda (z)$ for the 
optimally local, perfect staggered fermion
of mass 1, 2 and 4. The table contains all
the couplings $\geq 10^{-5}$.}
\end{table}

%\section{Spectral and thermodynamical properties of perfect staggered
\section{Spectral and thermodynamic properties of perfect staggered
fermions}

Now we want to address the question, in which sense the action derived
in section 2 is perfect, i.e. which observables are
free of artifacts due to the finite lattice spacing. 

First we look at the spectrum. For momentum $p=(\vec p ,p_{4})$
we see that the propagator has a pole at
$p_{4} = i \sqrt{\vec p^{\, 2} +m^{2}}$,
which corresponds to the {\em exact} continuum spectrum.
We note that the $\Pi$ function and the smearing term
do not affect the singularity structure of the propagator.
However, there are more poles --- one for each $\vec l$ --- namely
\footnote{Note that $\int_{-\pi}^{\pi} dp_{4}$ and the sum over $l_{4}$
combine to $\int_{-\infty}^{\infty} dp_{4}$.}
\begin{equation}
p_{4, \vec l} = i \sqrt{(\vec p + 2\pi \vec l )^{2} 
+ m^{2}} \ .
\end{equation}
These poles correspond to higher branches. 
%(we recall that
%$p_{\mu} \in ]-\pi ,\pi ]$). 
\footnote{Sometimes higher branches are referred to as
``ghosts'', which should not be confused, however,
with Faddeev-Popov ghosts.}
Additional branches are necessary for perfection, 
since the lattice imposes $2\pi $ periodicity.\\

As a test case for truncated perfect fermions, it is interesting
to observe how much harm we do to the spectrum -- in particular
to its lowest branch -- if we restrict the couplings
to a short range.
\begin{figure}[hbt]
\hspace{7mm}
\def\fpsangle{0}
\epsfxsize=120mm
\fpsbox{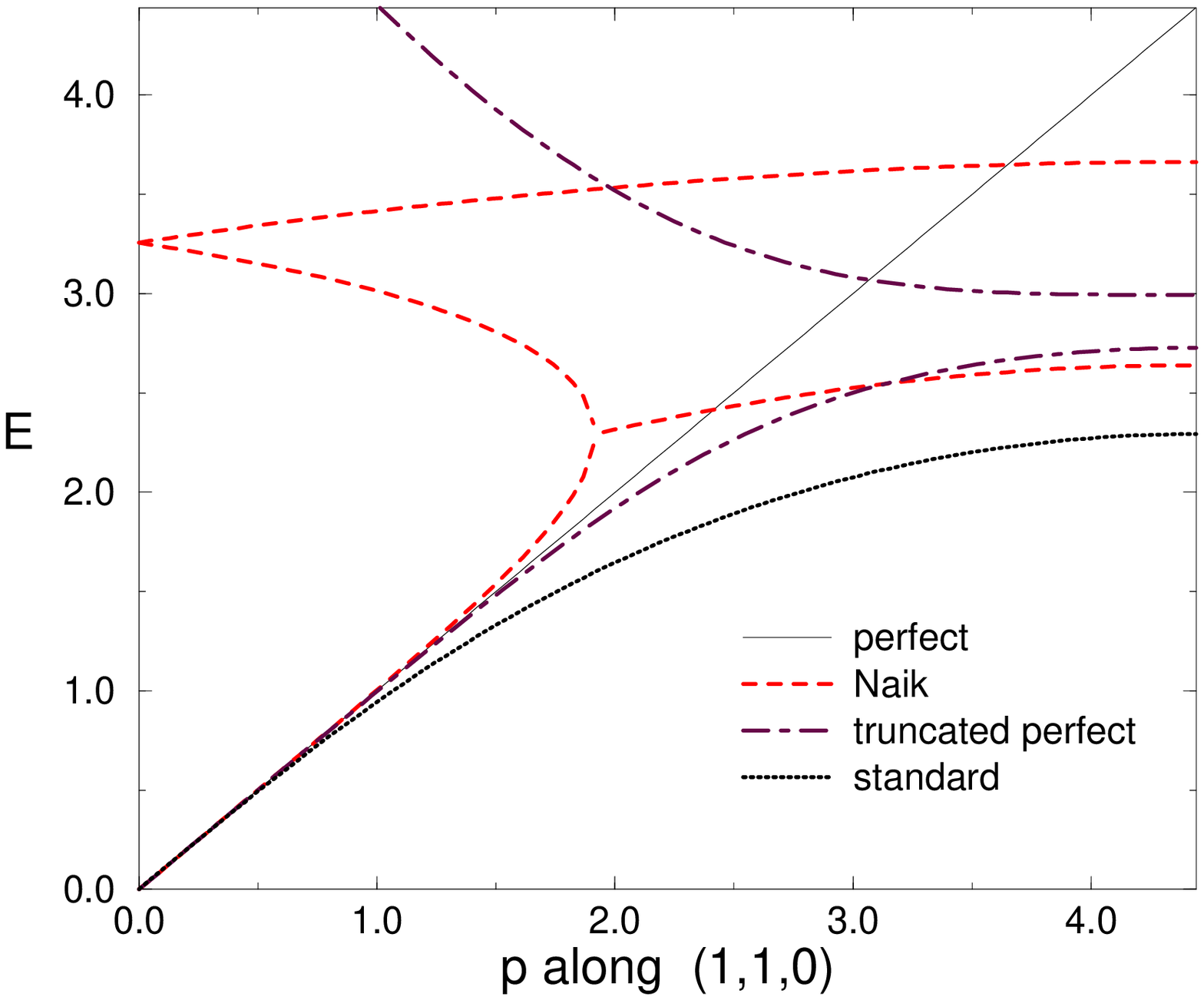}
\caption{Dispersion relation for truncated perfect, Naik
and standard, staggered fermions along (1,1,0).}
\vspace{-2mm}
\end{figure}
In Fig.~3 we show the spectrum for a truncated perfect
fermion, where we omit all couplings beyond $\pm 1/2, \pm 3/2$
in the $\mu $ direction for $\rho_{\mu}$, and all couplings
beyond $0,\pm 1$ in the non-$\mu$ directions of $\rho_{\mu}$
and in all directions of $\lambda$. The lower branch is
real and approximates the exact dispersion
relation better than the standard staggered fermion action,
and the upper branch is the real part of two complex
conjugate poles of the propagator.

An on-shell improved staggered fermion was proposed by 
S. Naik \cite{Naik}. By changing the kinetic
term of the standard action (\ref{standa}) to
\begin{equation} \label{snai}
\rho_{\mu ,z} = \Big( \frac{9}{8} [ \delta_{2z_{\mu},1}
- \delta_{2z_{\mu},-1}] - \frac{1}{24}
[ \delta_{2z_{\mu},3} - \delta_{2z_{\mu},-3}] \Big)
\prod_{\nu \neq \mu} \delta_{z_{\nu},0} \ ,
\end{equation}
he removed the $O(a^{2})$ artifacts for the free fermion. 
These couplings have no similarity to our perfect fermion.
For comparison, the resulting dispersion relation is also
shown in Fig.~3.
%% (Sch) I don't think that this is needed, and it really sounds
%% strange to me !
%and is denoted as ``Naik''. 
By construction, 
% that (Sch)
this dispersion relation
is good at $\vert p \vert << 1$, but at $\vert p_{\mu} \vert \sim 1$
the lower branch is hit by an upper branch, and then they turn
into two complex conjugate poles, the real part of which is shown
in the figure.

With respect to certain other quantities, even our
untruncated action is not exactly perfect.
When we performed the Gaussian integrals in the
RGT (\ref{RGT}), we did not keep track
of ``constant factors'', which do not depend on the
lattice fields. However, such factors may depend
on other quantities such as the temperature. If those
quantities are important -- which was not the case in 
the spectrum -- we can not expect perfect observables.
Then one notices that the lattice fields are not
completely renormalized. The possibility to keep track
of such factors in a block spin RGT was explored for the Ising
model in Ref. \cite{NauNie}.

As an example we consider the pressure of free, massless
fermions in infinite volume. In the continuum the relation
\begin{equation}
% P = \frac{7 \pi^{2}}{180} \cdot T^{4} 
% = 0.3838\dots \cdot T^{4} (Uwe,Sch)
\frac{P}{T^{4}} = \frac{7 \pi^{2}}{180}
\simeq 0.3838
\end{equation}
is known, where $P$ is the pressure and $T$ the temperature
(Stefan-Boltzmann law).
In Fig.~4 we plot the ratio $P/T^{4}$ for the staggered FPA
and for the Naik and standard and staggered action at 
$N_{t}=2,4,6 \dots 40$ lattice points in the 4-direction
($N_{t}$ must be even to accommodate a complete set of 
pseudoflavors).
The standard staggered action scales better
than Wilson fermions, in agreement with the fact that their
artifacts are of the second order in the lattice spacing,
whereas the Wilson action is plagued by linear artifacts~\footnote{
For Wilson fermions the ratio $P/T^{4}$ has a peak
% up to $\sim 1.9$ at small $N_{t}$ \cite{StL}. (Sch)
$\sim 1.9$ at $N_{t}\sim2,3$ \cite{StL}.
For a discussion of the artifacts in standard
staggered fermions and an improvement program for its
matrix elements, see \cite{Luo}.}.
We also see that even for the FPA this scaling
quantity differs from the exact continuum result. This
deviation is due to a missing temperature dependent
renormalization factor. However, the figure shows that
this factor is typically very close to 1.
It differs significantly only on immensely coarse lattices.
Hence the scaling is considerably improved with respect to
the standard lattice actions. This is still true after
truncation (the same we used when discussing the spectrum),
although the behavior gets somewhat worse. Truncation causes
an overshoot, which indicates that the fermion moves closer
to the standard formulation.
For the Naik fermion we confirm
some improvement as well. Its thermodynamic behavior has also 
been discussed in Ref.~\cite{Karsch}. The dip in the beginning
is very similar to the behavior of the D234 action
\cite{D234} (the corresponding thermodynamic plot is given in
\cite{StL}). Note that also the construction of those two
on-shell improved fermions -- Naik for the staggered and D234 for
the Wilson type fermions -- is very similar: in both cases additional 
couplings are added on the axes (which is not too promising for
the restoration of rotational invariance). Finally, the spectrum is
very similar too, the lowest branches are hit by a doubler and
turn complex around $\vert p_{\mu} \vert \sim 1$.

\begin{figure}[hbt]
\hspace{7mm}
\def\fpsangle{0}
\epsfxsize=120mm
\fpsbox{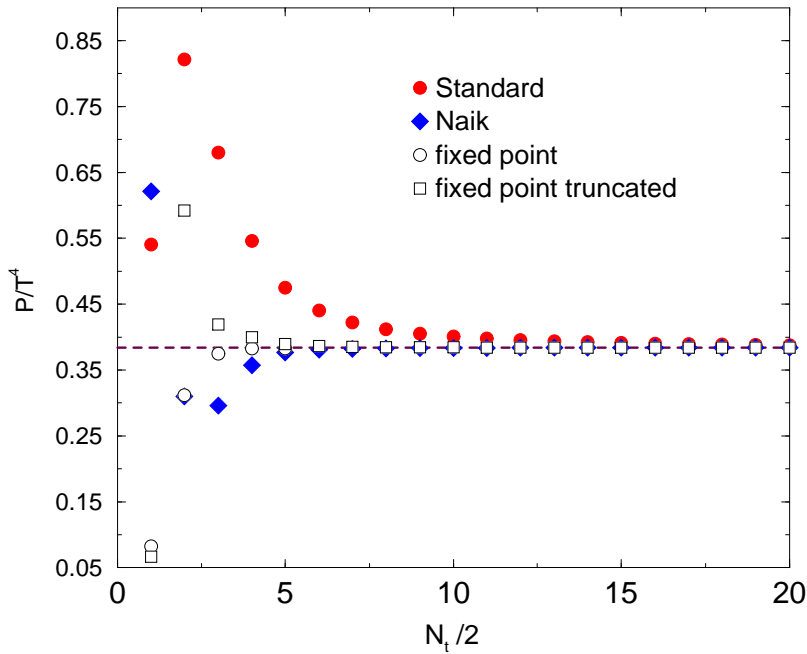}
\caption{The thermodynamic scaling quantity $P /T^{4}$
for free massless staggered fermions with the standard
action, the fixed point actions with and without
truncation, and the Naik action.}
\vspace{-2mm}
\end{figure}

%\section{A staggered fixed point action for free gauge fields} (Sch)
\section{A free fixed point gauge action consistent with 
staggered fermions}

Perfect staggered fermions can only be used in QCD if we
are able to couple them to fixed point gauge fields in
a perfect way. This requires a blocking scheme for the
gauge field, which is consistent with the blocking
for staggered fermions. The consistency condition
is based on gauge covariance.

Let's go back to a finite blocking factor $n$.
When we block from a fine to a coarse lattice, it is
convenient to fix a gauge for the fine lattice fields.
However, this gauge fixing should be restricted to one block,
in order to avoid long distance gauge dependence.

Following Ref.~\cite{Kalk} we block such that all contributions
to a coarse lattice variable have the same pseudoflavor,
as it is illustrated in Fig.~5. 
If a coarse pseudoflavor lives on a site $x'$,
then we consider the hypercube $n^{d}$ with center $x'$,
and all the $n^{d}$ fine variables of the same
pseudoflavor contribute to that block variable.
Thus each fine variable contributes
to one coarse variable of the same pseudoflavor.
(In the previous section we considered the limit $n\to \infty$
of this RGT.)

A coarse gauge field $A'_{\mu ,x'}$ in terms of fine
fields $A_{\mu ,x}$, which is consistent with gauge
covariance, is
\begin{equation} \label{congau}
A_{\mu ,x' + n\hat \mu /2}' = \frac{b_{n}}{n^{d}}
\sum_{x \in x'}  \frac{1}{n} \sum_{j=0}^{n-1} A_{\mu ,x
+(2j+1)\hat \mu /2} ,
\end{equation}
where $b_{n}$ is a constant renormalization factor.
Its value is chosen such that we obtain a finite FPA for the
free gauge field. Essentially, $b_{n}$ neutralizes the
constant factor from rescaling $A_{\mu}'$, hence dimensional
reasons suggest $b_{n}=n^{d/2-1}$.

Here our fine lattice has spacing 1/2 and the 
coarse one $n/2$, ($n$ odd). The sum $x \in x'$ runs over all 
$n^{d}$ fine lattice points that contribute to the coarse
pseudoflavor living at the site $x'$, and the lattice
gauge fields live on the link centers.
We sum over the straight connections of corresponding
fine lattice pseudoflavors contributing to adjacent
coarse lattice variables.
This construction is illustrated for $n=3$ in Fig.~5.
\begin{figure}[hbt]
\hspace{7mm}
\def\fpsangle{0}
\epsfxsize=120mm
\fpsbox{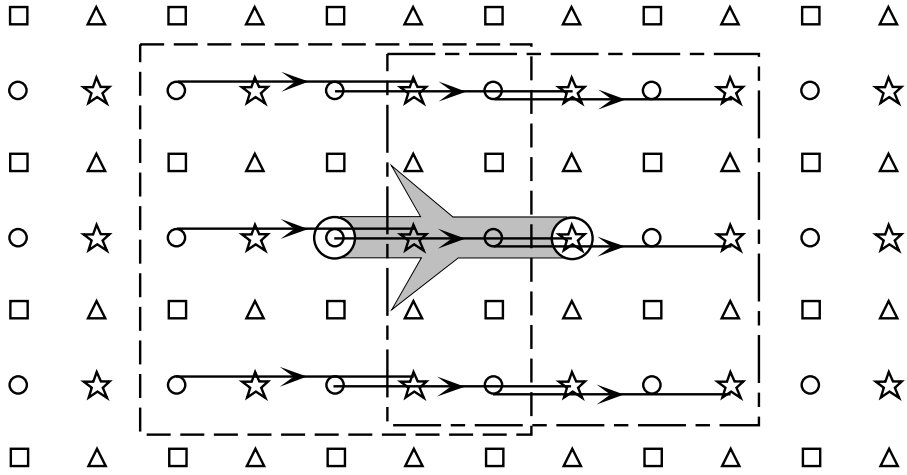}
\caption{The staggered blocking of fermions and gauge fields,
which maintains the pseudoflavor structure and gauge
covariance. This figure illustrates the case $d=2, \ n=3$.
Each symbol represents a pseudoflavor. Block variables
are encircled and pick up contributions of the fine variables
(with the same flavor) in the dashed box around them.
The block link (shaded) is built from all the links
covered by arrows.}
\vspace{-2mm}
\end{figure}

In momentum space this condition reads
\begin{eqnarray} \label{congaumo}
A'_{\mu}(p) &=& \frac{b_{n}}{n^{d}} \sum_{l'} A_{\mu}
(p+4\pi l'/n) \Pi^{M}_{n \mu} (p+4 \pi l'/n)(-1)^{l{'}_{\mu}} ,
\nonumber \\
\Pi^{M}_{n \mu} &=& \frac{\Pi^{M}_{\mu}(np)}{\Pi^{M}_{\mu}(p)} \ ,
\quad \Pi^{M}_{\mu}(p) = \frac{4 \sin (p_{\mu}/4)}{p_{\mu}}
\Pi (p) ,
\end{eqnarray}
where the summation extends over $l' \in \{ 0 ,1 ,2, \dots
,n-1\}^{d} $.
Note that the momenta of the fine fields are in the zone
$B_{4\pi} = ]-2\pi ,2\pi ]^{d}$, and those of the coarse fields in
$B_{4\pi /n} = ]-2\pi /n,2\pi /n]^{d}$.
Since we defined the gauge variables on the link centers,
$A_{\mu}(p)$ is $4 \pi $ antiperiodic in $p_{\mu}$.\\

Also here we want to send the blocking factor to infinity and
block from the continuum. We start from a space-filling set of 
pseudoflavors, together with a continuum
gauge field $a_{\mu}$. In the limit $n \to \infty $
the above condition (\ref{congau})
turns into a Riemann integral of the form
\begin{eqnarray}
A_{\mu ,x} &=& \int d^{d}y M_{\mu}(y)a_{\mu}(x-y) \nonumber \\
M_{\mu}(y) &=& \Big\{ \begin{array}{ccc}
M(y_{\mu}) && \vert y_{\nu} \vert \leq 1/2, \ \nu \neq \mu \\
0 && {\rm otherwise} \end{array} \nonumber \\
M(y_{\mu}) &=& \left\{ \begin{array}{ccc}
1 && \vert y_{\mu} \vert \leq 1/4 \\
3/2 - 2 \vert y_{\mu} \vert && 1/4 \leq \vert y_{\mu} \vert \leq 3/4 \\
0 && {\rm otherwise} \end{array} \right. \ .
\end{eqnarray}
This convolution is illustrated in Fig.~6. It is manifestly gauge
covariant.
Due to its architecturally interesting
shape we call the function $M_{\mu}$ the {\em mansard} function.
\begin{figure}[hbt]
\hspace{7mm}
\def\fpsangle{0}
\epsfxsize=120mm
\fpsbox{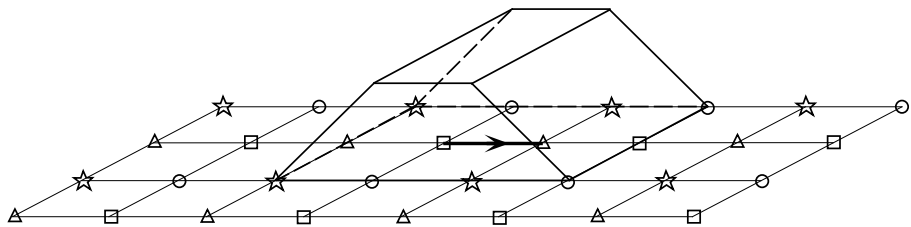}
\caption{The ``mansard blocking'' of a gauge field from the continuum.}
\vspace{-2mm}
\end{figure}

Note that
\begin{equation}
\int d^{d}y M_{\mu}(y) \exp (ipy) = \Pi^{M}_{\mu}(p) ,
\end{equation}
and therefore condition (\ref{congaumo}) turns into
\begin{equation} \label{congaucont}
A_{\mu}(p) = \sum_{l \in \Z^{d}} a_{\mu}(p+4\pi l)
\Pi_{\mu}^{M}(p+4\pi l)(-1)^{l_{\mu}} .
\end{equation}
The following property can be understood from the above
derivation, which started from a discrete RGT:
if we perform a gauge transformation on an Abelian
gauge field, $a_{\mu}(y) \to a_{\mu}(y)+ \partial_{\mu}
\varphi(y)$, then the lattice gauge field transforms as
$A_{\mu ,x} \to A_{\mu ,x} + \Phi_{x-\hat \mu /2}
- \Phi_{x+\hat \mu /2}$, where $\Phi_{x} = \int_{c_{x}} 
d^{d}x \varphi (x)$ 
($c_{x}$ being a unit hypercube with center $x$).
This shows gauge covariance and consistency with the fermionic
blocking. The same properties were achieved in 
Refs.\cite{Schwing,QuaGlu} for fermions of the Wilson type, 
where the blocking was done by stepwise integration
% summation resp. integration (Rich,Sch)
% I dont undertsand what this means, even Rich did not! Uwe seems
% to have understood it! It would be better to rewrite this in 
% a simpler way, perhaps just say:
over disjoint cells and the gauge field was convoluted with
a roof shaped function instead of the mansard function.

Now we construct the FPA, $S[A_{\mu}]$, for free gauge fields
, which is adequate for staggered fermions. We impose the Landau
gauge on the continuum gauge fields and choose the RGT
\begin{eqnarray}
\exp \{ -S[A_{\mu}] \} &=& \int \D a \D D 
\exp \Big\{ -\frac{1}{(2\pi )^{d}}
\int d^{d}p \frac{1}{2} a_{\mu}(-p) p^{2} a_{\mu}(p) \Big\}
\nonumber \\ & \times & \!\!
\exp \Big\{ -\frac{1}{(2\pi )^{d}} \int_{B_{4\pi}}
d^{d}p \frac{1}{2} D_{\mu}(-p) [ \alpha (p) + \gamma (p)
\widehat {(p_{\mu}/2)} ^{2} ] D_{\mu}(p) \nonumber \\
&& \hspace{-10mm} + i D_{\mu}(-p) [A_{\mu}(p)- \sum_{l \in \Z^{d}}
a_{\mu}(p+4\pi l) \Pi^{M}_{\mu}(p+4\pi l) (-1)^{l_{\mu}} ]
\Big\} , \label{gauRGT}
\end{eqnarray}
where the measure $\D a$ contains the
gauge fixing factor $\delta (\sum_{\mu}p_{\mu}a_{\mu}(p))$
% (Sch)
%. Therefore the continuum gauge propagator is simply $1/p^{2}$.
%\footnote{Actually this continuum propagator corresponds to the
%Feynman gauge, rather than the Landau gauge. However, in this
%context the use of $1/p^{2}$ is justified because in the
%FPA the difference vanishes.}
% ???
%% \footnote{In the derivation of the FPA we only solve classical
%% equations of motion. Hence we don't need the
%% Landau gauge propagator of the continuum action.}.
% ???  How about this version ?
Again we impose the blocking condition (\ref{congaucont})
only by a Gaussian
involving a mass-like and a kinetic smearing term.
$D_{\mu}$ is an auxiliary lattice field, defined on the same
links as $A_{\mu}$, in analogy to the fermionic
auxiliary fields $\bar \eta , \eta$ in eq.~(\ref{RGT}).
Performing the RGT (\ref{gauRGT}) yields
\begin{eqnarray}
S[A_{\mu}] &=& \frac{1}{(2\pi )^{d}} \int_{B_{4\pi}} d^{d}p
\frac{1}{2} A_{\mu}(-p) \Delta_{\mu}^{g}(p)^{-1}A_{\mu}(p)
\nonumber \\
\Delta_{\mu}^{g}(p) &=& \sum_{l \in \Z^{d}}
\frac{1}{(p+4\pi l)^{2}} \Pi^{M}_{\mu}(p+4\pi l)^{2}
+ \alpha (p) + \gamma (p) \widehat {(p_{\mu}/2)} ^{2} .
\end{eqnarray}
This is the FPA in a special gauge that we call
``fixed point lattice Landau gauge'',
where $A_{\mu}$ obeys
\begin{equation}
\sum_{\mu} \sin \frac{p_{\mu}}{4} \Delta_{\mu}^{g}(p)^{-1}
A_{\mu}(p) = 0 .
\end{equation}
We follow the standard procedure to rewrite the FPA in a
gauge invariant form and arrive at
\begin{eqnarray}
S[A] &=& \frac{1}{(2\pi )^{d}} \int_{B_{4\pi}} d^{d}p
\frac{1}{2} A_{\mu}(-p) \Delta_{\mu \nu}^{g}(p)^{-1}A_{\nu}(p)
, \nonumber \\ \label{FPAg}
\Delta_{\mu \nu}^{g}(p)^{-1} &=&
\Delta_{\mu}^{g}(p)^{-1} \delta_{\mu \nu}
- \frac{\sin(p_{\mu}/4) \Delta^{g}_{\mu}(p)^{-1} \Delta^{g}
_{\nu}(p)^{-1} \sin (p_{\nu}/4)}
{\sum_{\lambda} \sin^{2} (p_{\lambda}/4) 
\Delta^{g}_{\lambda}(p)^{-1}}.
\end{eqnarray} 
This result is similar to the one obtained for the
roof-like blocking in \cite{Schwing,QuaGlu}.
There it was possible to choose the smearing parameters such
that the FPA in $d=2$ turned into the ultralocal plaquette
action.
This can also be achieved for the ``staggered FPA'' given in
eq.~(\ref{FPAg}).

In $d=2$, the standard lattice Landau gauge implies
\begin{equation}
A_{\mu}(p) = i \epsilon_{\mu \nu} \frac{\sin (p_{\nu}/4)F(p)}
{\sum _{\lambda} 4 \sin^{2}(p_{\lambda}/4)} \ ,
\end{equation}
where $F$ is the plaquette variable defined on the plaquette centers.
Inserting this into eq.~(\ref{FPAg}) for $d=2$, we find
\begin{eqnarray}
S[F] &=& \frac{1}{(2\pi )^{2}} \int_{B_{4\pi}} d^{2}p \frac{1}{2}
F(-p) \rho (p) F(p) , \nonumber \\
\rho(p)^{-1} &=& 16 [ \sin^{2} \frac{p_{1}}{4} \Delta^{g}_{2}(p) +
\sin^{2} \frac{p_{2}}{4} \Delta^{g}_{1}(p)] \nonumber \\
&=& \cos^{2}\frac{p_{1}}{4} \cos^{2}\frac{p_{2}}{4}
\Big( 1 - \frac{1}{6} \widehat {(p_{1}/2)} ^{2}) \Big)
\Big( 1 - \frac{1}{6} \widehat {(p_{2}/2)} ^{2}) \Big) \nonumber \\
&& + 4 \alpha (p) [ \widehat {(p_{1}/2)} ^{2} +
\widehat {(p_{2}/2)} ^{2} ] + 8 \gamma (p)
\widehat {(p_{1}/2)} ^{2} \widehat {(p_{2}/2)} ^{2} .
\end{eqnarray}
If we choose
\begin{eqnarray}
\alpha (p) &=& \frac{1}{16} +\frac{1}{24}
 \cos^{2}\frac{p_{1}}{4} \cos^{2}\frac{p_{2}}{4} \nonumber \\
\gamma (p) &=& - \frac{1}{128} - \frac{1}{288}
\cos^{2}\frac{p_{1}}{4} \cos^{2}\frac{p_{2}}{4} \label{gRGTpar} ,
\end{eqnarray}
then we obtain $\rho^{-1}(p) = 1$, as desired.
Taking this as a guide for higher dimensions suggests
that the optimal choice for the RGT parameters is
\begin{eqnarray}
\alpha (p) &=& \frac{1}{16} +\frac{1}{24}
\prod_{\nu =1}^{d} \cos^{2}\frac{p_{\nu}}{4} \nonumber \\
\gamma (p) &=& - \frac{1}{128} - \frac{1}{288}
\prod_{\nu =1}^{d} \cos^{2}\frac{p_{\nu}}{4} . \label{rgtparga}
\end{eqnarray}
It is important that $\alpha (p) + \gamma (p)
\widehat {(p_{\mu}/2)}^{2}$ is always positive. This ensures that the 
functional integrals in the RGT (\ref{gauRGT}) are well defined.

%It turns out that this RGT yields indeed a very local FPA in $d=4$. (Rich)
Indeed it turns out that this RGT yields a very local FPA in $d=4$.
The largest couplings are given in table 3 and the exponential decay
is plotted in Fig.~7.
\begin{table}
\begin{center}
\begin{tabular}{|c|c|c|c|c|}
\hline
$(z_{1},z_{2},z_{3},z_{4})$ & $\rho_{11}(z)$ && $(z_{1},z_{2},z_{3},z_{4})$ 
& $\rho_{12}(z)$ \\
\hline
(0,0,0,0) & ~3.8112883 && (1/2,1/2,0,0) & ~0.6609129 \\
\hline
(0,0,0,1) & -0.3402240 && (1/2,1/2,0,1) & ~0.0742258 \\
\hline
(1,0,0,0) & -0.1541891 && (1/2,3/2,0,0) & -0.0271129 \\
\hline
(0,0,1,1) & -0.1150765 && (1/2,1/2,1,1) & ~0.0148817 \\
\hline
(1,0,0,1) & ~0.0508985 && (1/2,3/2,0,1) & ~0.0047334 \\
\hline
(0,1,1,1) & -0.0476001 && (3/2,3/2,0,0) & ~0.0088456 \\
\hline
(1,0,1,1) & -0.0058551 && (1/2,3/2,1,1) & ~0.0015899 \\
\hline
(1,1,1,1) & -0.0077244 && (3/2,3/2,0,1) & -0.0003598 \\
\hline
(0,0,0,2) & ~0.0054232 && (3/2,3/2,1,1) & -0.0010820 \\
\hline
(2,0,0,0) & ~0.0084881 && (1/2,1/2,0,2) & -0.0041648 \\
\hline
(0,0,1,2) & -0.0012443 && (1/2,5/2,0,0) & ~0.0015083 \\
\hline
(1,0,0,2) & -0.0065388 && (1/2,1/2,1,2) & -0.0001960 \\
\hline
(2,0,0,1) & -0.0039935 && (1/2,3/2,0,2) & ~0.0004785 \\
\hline
(0,1,1,2) & -0.0006582 && (1/2,5/2,0,1) & -0.0005171 \\
\hline
(2,0,1,1) & ~0.0011516 && (3/2,5/2,0,0) & -0.0007231 \\
\hline
(1,1,1,2) & ~0.0004326 && (3/2,3/2,0,2) & -0.0002568 \\
\hline
(2,0,0,2) & ~0.0011159 && (3/2,5/2,0,1) & ~0.0001332 \\
\hline
(1,0,2,2) & ~0.0005143 && (3/2,3/2,1,2) & -0.0002152 \\
\hline
(1,1,2,2) & ~0.0003562 && (1/2,1/2,2,2) & ~0.0001109 \\
\hline
(1,2,1,1) & ~0.0002912 && (5/2,5/2,0,0) & ~0.0001907 \\
\hline
(0,0,0,3) & -0.0005163 && (1/2,1/2,0,3) & ~0.0002217 \\
\hline
(3,0,0,0) & -0.0005617 && (1/2,7/2,0,0) & ~0.0001005 \\
\hline
(3,0,0,1) & ~0.0003064 && (1/2,3/2,0,3) & -0.0001036 \\
\hline
\end{tabular}
\end{center}
\caption{The largest couplings for the fixed point action of the
free gluon with respect to the ``mansard RGT''.
The table includes all couplings in $\rho_{11}$ with values
$\geq 2.5 \cdot 10^{-4}$ and all couplings in
$\rho_{12}$ with values $\geq 10^{-4}$.}
\end{table}
\begin{figure}[hbt]
\hspace{7mm}
\def\fpsangle{0}
\epsfxsize=120mm
\fpsbox{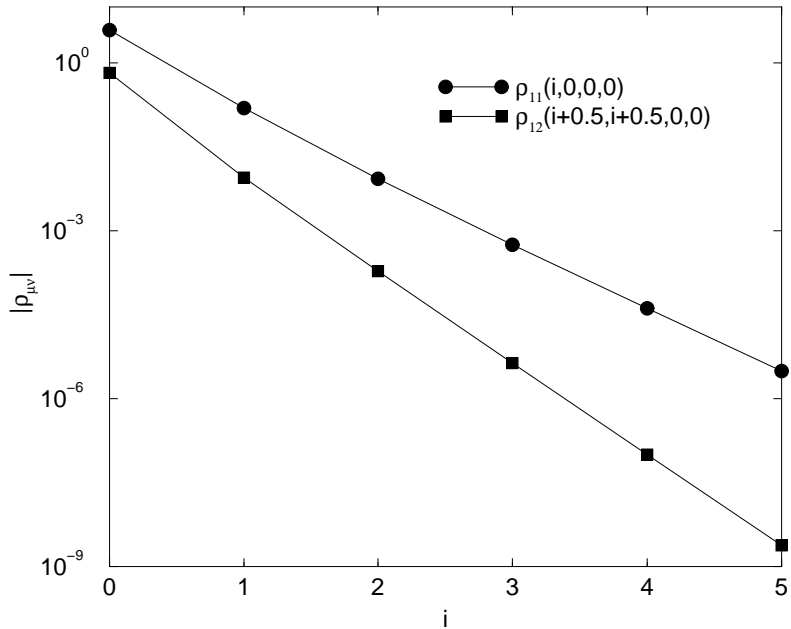}
\caption{The exponential decay of the free gluon FPA for a RGT 
of the mansard type.}
\vspace{-2mm}
\end{figure}

% Uwe is uncertain about the ``perfect'' gluon having only the
% lowest branch perfect. I think now I understand what you mean, I will
% try to talk to him about this.
The dispersion relation of the transverse mansard gluon is
again perfect, as it is the case
for the perfect free fermion.
If we truncate the couplings
to distances $\leq 3/2$, we obtain the dispersion shown in Fig.~8.
Due to our optimization of locality, the spectrum of the
truncated fixed point gluon approximates the continuum
spectrum still very well for small and moderate momenta. 
% (Sch)
In practice, since these gauge fields are designed to couple to
staggered fermions, the relevant momentum region is $|p_{\mu}| < \pi$,
where
the dispersion is quite good. We will find further evidence for this 
from the calculation of the quark-antiquark potential.
\begin{figure}[hbt]
\hspace{7mm}
\def\fpsangle{0}
\epsfxsize=120mm
\fpsbox{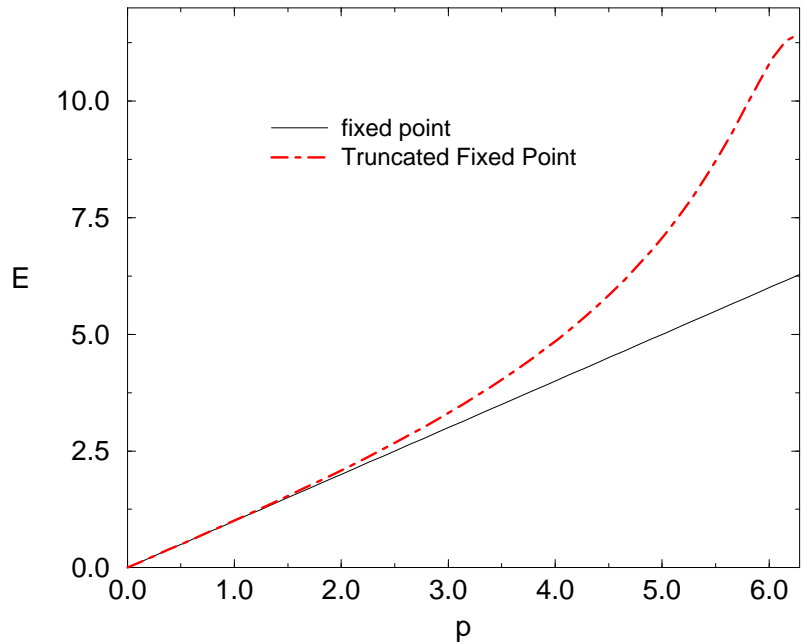}
\caption{The dispersion relation of the 
truncated mansard fixed point gluon.}
\vspace{-2mm}
\end{figure}

% The following section gets an error in the latex? overfull hbox,
% can you resolve this? (Sch)
\section{Polyakov loop and the static quark-antiquark potential}

It is also possible to construct perfect operators for a given RGT.
One adds a source term to the continuum action (or the fine lattice action)
and includes this term perturbatively in the blocking process.
In Ref.~\cite{GN} the perfect $\bar \chi \chi$ operator was
constructed in this way. In principle this method allows
% (Rich)
us to build any perfect composite operators, but in practice
this tends to be difficult.

It is easier to construct ``classically perfect'' fields
and operators. Classically perfect fields are obtained by 
minimizing the
continuum action together with the blocking transformation term.
If one inserts $\hbar$ in the RGT expression (\ref{gauRGT}), 
then one obtains them in the limit $\hbar \to 0$.
%where to functional integrals
%do not have to be carried out any more.

% Uwe suggests using large bars intsead of -- below. I have just added one 
% more - at each!
For the free gauge field --- and the
RGT considered above --- the classically perfect field reads
\begin{equation} \label{claga}
a^{c}_{\mu}(p) = \frac{1}{p^{2}} \Pi^{M}_{\mu}(p)
\Delta^{g}_{\mu}(p)^{-1} A_{\mu}(p) ,
\end{equation}
where $a_{\mu}^{c}$ is in the Landau gauge and
$A_{\mu}$ in the fixed point lattice Landau gauge.
Note that classically perfect fields are defined in the continuum.

We obtain ``classically perfect operators'' from continuum
operators if we replace the continuum fields by classically
perfect fields, which are then expressed in terms of lattice 
fields. For instance, we can build a classically
perfect Polyakov loop as
\begin{equation}
\phi^{c}(\vec x ) = \int dx_{d} \ a_{d}^{c}(\vec x , x_{d}).
\end{equation}
Inserting the gauge field given in eq.~(\ref{claga}), we obtain in
momentum space
\begin{equation}
\phi^{c}(\vec p ) = \frac{1}{\vec p^{\, 2}} \Pi^{M}_{d}(\vec p ,0)
\Delta_{d}^{g}(\vec p ,0)^{-1} A_{d}(\vec p ,0) .
\end{equation}
Also this operator is defined in the continuum, in contrast
to the standard lattice Polyakov loop 
$\Phi (\vec p )=A_{d}(\vec p ,0)$.
Of course we can restrict the classically perfect Polyakov loop
to lattice points by imposing $2\pi$ periodicity,
\begin{equation}
\Phi^{c}(\vec p ) = \sum_{\vec l \in \Z^{d-1}}
\frac{1}{(\vec p + 2\pi \vec l)^{2}} \Pi^{M}_{d}
(\vec p + 2\pi \vec l,0) \Delta^{g}_{d}(\vec p,0)^{-1}
A_{d}(\vec p,0).
\end{equation}

In $d=2$ we found $\Delta_{2}^{g}(p_{1},0)^{-1}=\hat p^{2}_{1}$,
which leads to
\begin{equation}
\Phi^{c}_{x} = \frac{1}{4} ( \Phi_{x+1}+2\Phi_{x}+\Phi_{x-1} ),
\end{equation}
i.e. the classically perfect Polyakov loop is ultralocal
on the 2d lattice. This confirms that the RGT parameters
introduced in Sec. 4 optimize locality.

The correlation function of two classically perfect Polyakov loops,
\begin{equation}
\langle \phi^{c}(-\vec p ) \phi^{c}(\vec p ) \rangle =
\frac{1}{(\vec p^{\, 2})^{2}} \Pi^{M}_{d}(\vec p ,0)^{2}
\Delta_{d}^{g}(\vec p,0)^{-1}
\end{equation}
yields the static quark-antiquark potential
\begin{equation} \label{statpot}
V(\vec r ) = - \frac{1}{(2\pi )^{d-1}} \int d^{d-1}p
\frac{1}{( \vec p ^{\, 2})^{2}} \Pi^{M}_{d}(\vec p ,0)^{2}
\Delta_{d}^{g}(\vec p ,0)^{-1} \exp (i\vec p \, \vec r) .
\end{equation}
In $d=2$ this integral diverges as it stands.
We subtract an infinite constant such that 
we obtain the correct behavior at large 
$r$. Then integration by contour techniques leads to
\begin{equation}
% V_{d=2}(r) = \left\{ \begin{array}{ccc} (Uwe)
V(r) = \left\{ \begin{array}{cc}
%r/2 && r \geq 1.5 \\
%(-r^{5}+15r^{4}-90r^{3}+270r^{2}-165r+243)/480 && 1 \leq r \leq 1.5 \\
%(r^{5}-5r^{4}-10r^{3}+110r^{2}-5r+179)/480 && 0.5 \leq r \leq 1 \\
%(2r^{5}-10r^{4}+100r^{2}+178)/480 && 0 \leq r \leq 0.5 \end{array} 
r/2 & r \geq 1.5 \\
\!\!\!\! (-32r^{5}+240r^{4}-720r^{3} 
+1080r^{2}-330r+243)/960 & 1 \leq r \leq 1.5 \\
(32r^{5}-80r^{4}-80r^{3}+440r^{2}-10r+179)/960 & 0.5 \leq r \leq 1 \\
(64r^{5}-160r^{4}+400r^{2}+178)/960 & 0 \leq r \leq 0.5 \end{array} 
\right. 
\end{equation}
This potential is shown in Fig.~9. It is very smooth (four times 
continuously differentiable) and coincides with the exact potential
at $r\geq 1.5$.
\begin{figure}[hbt]
\hspace{7mm}
\def\fpsangle{0}
\epsfxsize=120mm
\fpsbox{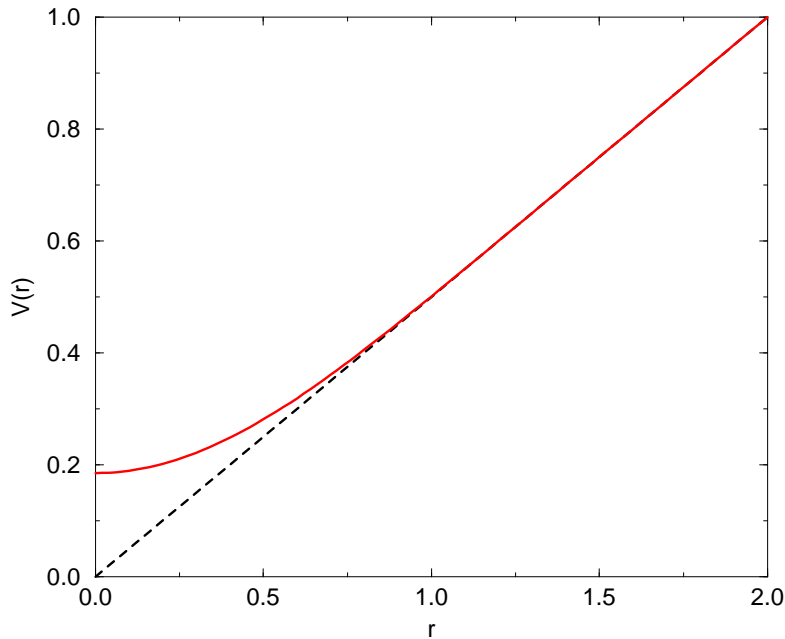}
\caption{The classically perfect static quark-antiquark potential in $d=2$.}
\vspace{-2mm}
\end{figure}

In $d=4$ the Fourier transform (\ref{statpot}) has to be calculated
numerically. The result is shown in Fig.~10 and compared to the
static potential based on the standard lattice Polyakov loop,
which is only defined if $\vec r$ is a lattice vector.
For the classically perfect potential,
we observe a faster convergence of the scaling quantity
%$\vert \vec r \vert \cdot V(\vec r)$ to the exact value $-1/4\pi $. 
% (Uwe,Sch)
$\vert \vec r \vert V(\vec r)$ to the exact value $-1/4\pi $.
Also rotational invariance is approximated much better,
even down to $ \vert \vec r \vert <1$.
The reason for the remaining artifacts in the classically perfect
potential is related to the reason for imperfectness of the pressure
discussed in Sec. 3. 
%Also here we (Rich) 
Here we also 
ignore ``constant factors''
in the Gaussian integrals of the RGT by taking just the minimum
of the exponent. The remaining artifacts are 
exponentially suppressed.
% Dont include the sentence below (as you have already done!) (Uwe, Sch)
%The potential corresponding to a truncated mansard gluon,
%on the other hand, is quite a disaster. ???
\begin{figure}[hbt]
\hspace{7mm}
\def\fpsangle{0}
\epsfxsize=120mm
\fpsbox{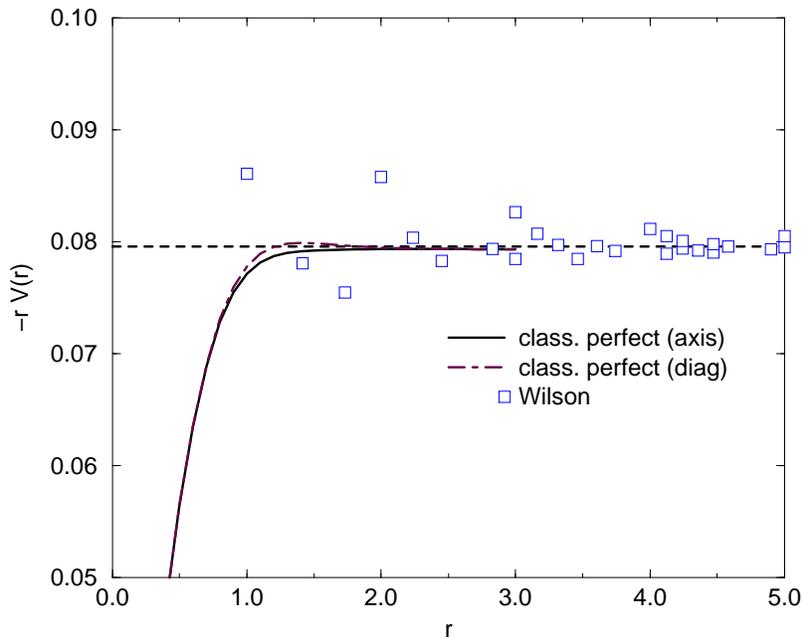}
\caption{The classically perfect static quark-antiquark potential 
in $d=4$ on an axis and on the space diagonal, compared to the
potential obtained from Wilson's standard plaquette action.}
\vspace{-2mm}
\end{figure}

\section{Conclusions and outlook}

We have constructed perfect free staggered fermions and perfect
free gauge fields, which can be coupled consistently. This
forms a basis for systematic perfect lattice perturbation
theory, leading to a perturbatively perfect lattice action
for QCD with staggered fermions.
The next step is the evaluation of the perfect quark-gluon
% vertex function, 
% (Sch) Should we add the three gluon vertex? Makes it systematic!
and the three gluon vertex functions 
as it was done for Wilson-like fermions \cite{QuaGlu,StL,prep}. In momentum 
space these vertex functions have almost the same form as written down in
\cite{QuaGlu}, with some modifications for staggered fermions
and mansard gluons, which are obvious from the present paper.
% (Sch)
%This action may be applicable to simulations of heavy
% quarks. 
% There the worst artifacts of the standard lattice
% formulation are amplified by the mass and remarkably suppressed
% for actions, which are perfect at weak couplings.
%\{to be justified better ???\}
% (Uwe suggested that it seems somewhat odd that we are suggesting
% heavy quark simulations for staggered fermions.)
The lattice action thus obtained would be perfect for weak couplings. It 
may already be useful in lattice simulations even at moderate
correlation length,
since the classically perfect action is also expected to be one-loop perfect. 
The next step is to test the actions in simple situations.
For example, one could study heavy quark physics, even though staggered 
fermions are not designed for such applications. In such a study, the 
worst artifacts of the standard lattice formulations originate from the large 
mass of the quarks. However, they are considerably suppressed for actions 
which are perfect at weak couplings.
% We observed for Wilson fermions  (Sch)
For Wilson fermions, we observed 
that even the use of
the free perfect fermion action, together with standard
lattice gauge fields, improves the 
% scaling drastically (Sch; We did not study any scaling?)
% the word drastic sounds negetive to my ears!
mesonic dispersion dramatically
\cite{StL} --- although a large additive mass renormalization 
% takes place. (Uwe)
occurs.
%(Sch)
For staggered fermions, only a multiplicative mass renormalization
will occur due to chiral symmetry.

% I switched the next two paragraphs since it seemed natural to discuss
% the applications of the present work before talking about the need
% to do non-perturbative work. (Sch)

An important aspect of staggered fermions is that they break the
%(pseudo)- (Uwe) we dont have to mention pseudo!
flavor symmetry when interacting 
% by (Rich)
with gauge fields. In lattice QCD with staggered fermions
there is only one true ``Goldstone pion''. The other pions
remain massive in the chiral limit, which manifests the flavor
symmetry breaking \cite{pion}.
% (Sch)
This is a major obstacle in studying chiral symmetry breaking
at finite temperatures, since the number of flavors plays an important
role in such studies.
In an experiment with a ``fat link'' --- consisting of a link plus
staples with varying staple weight --- it was possible to reduce
the mass of the remaining pions by about a factor two \cite{Blum},
which shows that there is a large potential for improvement
by using non standard actions. However, the use of the Naik
fermion does not seem to help here \cite{Bernard}.
When we couple perfect staggered
fermions and gauge fields consistently, we expect 
the flavor symmetry breaking to be strongly reduced. 
% (Sch)
%This could help to make realistic studies of chiral
%phase transitions at finite temperature feasible.
The ``fat link'' would naturally be incorporated.
This could help to make the studies of chiral
phase transitions at finite temperature more realistic.
% (Sch)
As a simpler experiment one could combine the truncated
perfect staggered fermion with a standard lattice gauge field.
Perhaps this already helps to decrease the flavor symmetry breaking.
%pion mass splitting significantly. (Sch)

% To include also light quarks, 
% (Sch)
  In principle, if the dynamics of the problem involve large gauge fields,
as is obviously true for the physics of the light pions, it is unclear if
the improvements obtained perturbatively are sufficient at the typical
couplings where simulations can be performed. The couplings obtained
perturbatively could undergo further renormalizations and new couplings may
arise.\footnote{However, we would like
to emphasize that the perturbatively perfect action will approach the scaling
region must faster than standard actions.}
In such a situation one has to resort to non-perturbative techniques in order 
to obtain the classically perfect action. In this case 
one rescales the QCD action with the gauge coupling $g$ and looks for a fixed
point -- i.e. a classically perfect action --
of the rescaled theory. This can be done 
numerically by real space RGT steps, using the parameters for
RGTs with finite blocking factors given in the appendix.
The perturbatively
perfect action may serve as a promising point of departure for
this iteration. 
Thanks to the asymptotic freedom at $g=0$, such RGT steps
only consist of minimization of the action on the fine
lattice together with the transformation term.
% No (numeric) functional integral is needed, which simplifies (Uwe)
No (numerical) functional integration needs to be performed, which 
simplifies the task enormously. In practice one starts from 
% a random generated (Uwe) 
an updated configuration on the coarse lattice and determines
the minimizing fields on the fine lattice.
%This is the way to
%incorporate the effects of strongly fluctuating field configurations. (Sch)
In this way,
the effects of strongly fluctuating field configurations
can be incorporated.

%It is analogous to the experiment with mesons 
%built from perfect Wilson fermions 
%and standard lattice gauge fields, mentioned above.
%Due to the strong mass renormalization one would have
%to use a clearly negative bare mass in order to deal with
%pions. \\ (Uwe,Sch)
%Work along these lines is in progress. 

\vskip 0.5in
{\em Acknowledgment} \\

One of us, W.B., thanks F. Karsch for useful
comments.

\newpage
%\vskip0.5in
% (Rich)
\begin{center}
{\Large\bf Appendix}
\end{center}

\appendix 

\section{Parameters for RGTs with finite blocking factors
and the same fixed point}

In the limit $g=0$ (where we determine the FPA),
the quarks decouple from the gauge fields, hence we can start
by constructing a pure gauge FPA. Here we provide the analytic
ingredients for this purpose.

In Sec.4 we derived a FPA for free gauge field, which
is consistent with the gauge requirements of staggered
fermions. There we blocked from the continuum, but of course
the same FPA can also be obtained using finite blocking
factor RGTs. The optimal RGT parameters, which provide ultralocality
in $d=2$ and extreme locality in higher dimensions, depend
on this blocking factor $n$ ($n$ odd). So far we only gave
those parameters at $n \to \infty$. In this appendix we are
going to identify them for general $n$. This is needed
for the nonperturbative search of fixed points of
non-Abelian gauge fields. There
one does RGTs numerically, and is therefore restricted
to finite blocking factors. In practice one would
choose the smallest value $n=3$.

Assume we are at the fixed point of Sec.4 in the fixed point
lattice Landau gauge, and we perform a block factor $n$ RGT,
\begin{eqnarray}
\exp \{ -S' [A_{\mu}'] \} &=& \int \D A \D D 
\exp \{ -S[A_{\mu}] \} \nonumber \\
& \times & \exp \Big\{ - \Big( \frac{n}{2\pi } \Big)^{d}
\int_{B_{4\pi /n}} \!\!\!\! d^{d}p \ 
\frac{1}{2} D_{\mu}(-p) \Omega_{\mu ,n}(p)
D_{\mu}(-p) \nonumber \\ && \hspace{-36mm}
+ i D_{\mu}(-p) \Big[ A_{\mu}'(p) - \frac{b_{n}}{n^{d}}
\sum_{l'} A_{\mu}(p+4\pi l'/n) \Pi^{M}_{n\mu}(p+4\pi l' /n)
(-1)^{l_{\mu}'} \Big] \Big\} ,
\end{eqnarray}
where $l' \in \{ 0,1,2, \dots , n-1 \}^{d}$ as in condition
(\ref{congaumo}), which is (smoothly) implemented here.
For the smearing term $\Omega_{\mu ,n}$ we need
an ansatz, which involves more parameters than it was the case
for the blocking from the continuum,
% (Sch) I added a factor of 4 in the gamma and sigma pieces
\begin{eqnarray}
\Omega_{\mu ,n}(p) &=& \alpha_{n} + 4 \sin^{2}(np_{\mu}/4) \gamma_{n}
+ \sin^{2}(np_{\mu}/2) \omega_{n} \nonumber \\
&& + [ \delta_{n}+4 \sin^{2}(np_{\mu}/4) \sigma_{n} ] \prod_{\nu =1}^{d}
\cos^{2} (np_{\nu}/4) .
\end{eqnarray}
The RGT parameters $b_{n}, \ \alpha_{n}, \ \gamma_{n}, \ \omega_{n}, 
\ \delta_{n}$ and $\sigma_{n}$ can now be determined from the condition 
that the action be invariant under this RGT. Doing this integral and 
rescaling the coarse lattice momenta into the full zone
$B_{4\pi}$, we obtain
\begin{eqnarray}
S'[A_{\mu}'] &=& \exp \Big\{ - \frac{1}{(2\pi )^{d}} \int_{B_{4\pi}}
d^{d}p \frac{1}{2} A_{\mu}'(-p) \Delta_{\mu}^{g}{'}(p)^{-1} A_{\mu}'(p)
\Big\} , \nonumber \\
\Delta_{\mu}^{g}{'}(p) &=& \frac{b_{n}^{2}}{n^{d}}
\sum_{l'} \Delta^{g}_{\mu} ((p+4\pi l')/n) \Pi^{M}_{n \mu}
% ((p+4\pi l')/n)^{2} + \Omega_{\mu ,1}(p) . (Sch)
((p+4\pi l')/n)^{2} + \Omega_{\mu ,n}(p) .
\end{eqnarray}
We combine the sum over $4\pi l'/n$ with the sum over $4\pi l , \
l \in \Z^{d}$, which is intrinsic in $\Delta_{\mu}^{g}$,
to a sum over $4\pi \ell /n , \ \ell \in \Z^{d}$. 
After doing some lengthy algebra
and evaluating a number of trigonometric sums,
we find that the fixed point condition $\Delta_{\mu}^{g}{'}(p) =
\Delta_{\mu}^{g}(p)$ is fulfilled 
% iff (Lets not be too mathematical!) (Uwe,Sch)
if
\begin{eqnarray}
b_{n}^{2} &=& n^{d-2} \ , \nonumber \\
2 \alpha_{n} &=& 3 \delta_{n} = 8n^{2} \omega_{n} = \frac{n^{2}-1}
{8 n^{2}} \ , \nonumber \\
4 \gamma_{n} &=& 9 \sigma_{n} = - \frac{(n^{2}-1)^{2}}{32 n^{4}} \ .
\end{eqnarray}
We see that $\Omega_{\mu ,n}(p)$ is always positive, such that
$\int \D D$ is well-defined.
In particular, we confirm the result for $b_{n}$, 
which we anticipated
%justified (Uwe)
in Sec.4 by dimensional
reasons, and in the limit $n \to \infty $ we reproduce the
parameters for the blocking from the continuum obtained in
eq.~(\ref{gRGTpar}).
As a further check we notice that $n=1$ yields a trivial
identity transformation, as it should.

\end{document}